\documentclass[fleqn,usenatbib]{mnras}

\usepackage{newtxtext,newtxmath}

\usepackage[T1]{fontenc}
\usepackage{ae,aecompl}
\usepackage{multirow}

\usepackage{graphicx}	
\usepackage{amsmath}	
\usepackage{hyperref}
\usepackage{footmisc} 

\usepackage{booktabs}
\usepackage{xspace}
\usepackage{verbatim}

\newcommand{\apx}[1]{^{\rm #1}}
\newcommand{\pdx}[1]{_{\rm #1}}

\def\pdot {\dot P}

\def\edot {\dot E}
\def\nudot {\dot \nu}
\def\nudotdot {\ddot \nu}

\def\nh{$N_{\rm H}$\xspace}

\def\col{cm$^{-2}$\xspace}

\def\flux{erg~s$^{-1}$~cm$^{-2}$\xspace}
\def\lum{erg~s$^{-1}$\xspace}
\def\msun{~M_{\odot}}

\def\deg{^\circ}

\def\psrj{PSR J1740$+$1000\xspace}

\def\xmm{{\em XMM-Newton}\xspace}

\def\cha{{\em Chandra}\xspace}

\def\fermi{{\em Fermi}\xspace}
\usepackage{color}

\title[Thermal and non-thermal X-ray emission from \psrj]{Thermal and non-thermal X-ray emission from the rotation-powered radio/$\gamma$-ray pulsar \psrj}

\author[Rigoselli et al.]{Michela Rigoselli$^{1}\thanks{E-mail: michela.rigoselli@inaf.it}$, Sandro Mereghetti$^{1}$, Sara Anzuinelli$^{1,2}$, Michael Keith$^{3}$,
\newauthor
Roberto Taverna$^{4}$, Roberto Turolla$^{4,5}$, Silvia Zane$^{5}$\\
$^{1}$ INAF, Istituto di Astrofisica Spaziale e Fisica Cosmica Milano, via A.\ Corti 12, I-20133 Milano, Italy\\
$^{2}$ Dipartimento di Fisica G. Occhialini, Università degli Studi di Milano Bicocca, Piazza della Scienza 3, I-20126 Milano, Italy\\
$^{3}$ Jodrell Bank Centre for Astrophysics, Department of Physics and Astronomy, The University of Manchester, Manchester M13 9PL, UK\\
$^{4}$ Dipartimento di Fisica e Astronomia, Università di Padova, Via F. Marzolo 8, I-35131 Padova, Italy\\
$^{5}$ MSSL, University College London, Holmbury St. Mary, UK
}

\date{Accepted 2022 April 21. Received 2022 April 20; in original form 2022 April 4}
\pubyear{2022}

\begin{document}
\label{firstpage}
\pagerange{\pageref{firstpage}--\pageref{lastpage}}
\maketitle

\begin{abstract}
We report the results of new \xmm observations of the middle-aged ($\tau_c=1.1\times10^5$ yr) radio pulsar \psrj carried out in 2017-2018. These long pointings ($\sim$530 ks) show that the non-thermal emission, well described by a power-law spectrum with photon index $\Gamma=1.80 \pm 0.17$, is pulsed with a $\sim$30\% pulsed fraction above 2 keV. 
The thermal emission can be well fit with the sum of two blackbodies of temperatures $kT_1=70\pm4$ eV and $kT_2=137\pm7$ eV, and emitting radii $R_1=5.4_{-0.9}^{+1.3}$ km and $R_2=0.70_{-0.13}^{+0.15}$ km (for a distance of 1.2 kpc). We found no evidence for absorption lines as those observed in the shorter \xmm observations ($\sim$67 ks) of this pulsar carried out in 2006.
The X-ray thermal and non-thermal components peak in anti-phase and none of them is seen to coincide in phase with the radio pulse. This, coupled with the small difference in the emission radii of the two thermal components, disfavors an interpretation in which the dipolar polar cap is heated by magnetospheric backward accelerated particles. 
Comparison with the other thermally emitting isolated neutron stars with spectra well described by the sum of two components at different temperatures shows that the ratios $T_2$/$T_1$ and $R_2$/$R_1$ are similar for objects of different classes. The observed values cannot be reproduced with simple temperature distributions, such as those caused by a dipolar field, indicating the presence of more complicated thermal maps.
\end{abstract}


\section{Introduction}
The radio pulsar \psrj was discovered in 2000 in a survey performed with the Arecibo radiotelescope \citep{2002ApJ...564..333M}. Its timing parameters  ($P=0.154$ s,  $\pdot=2.15\times 10^{-14}$ s~s$^{-1}$) imply that \psrj is an ordinary rotation-powered pulsar (RPP) with characteristic age $\tau_c = 1.1\times 10^{5}$ yr,   
spin-down luminosity $\edot=2.3\times 10^{35}$ \lum, and dipolar magnetic field $B\pdx{s} = 1.8\times 10^{12}$ G at the neutron star surface. Its distance, inferred from the dispersion measure using the \citet{yao17}  electron-density model of the Galaxy, is about 1.2 kpc.

This pulsar has a particularly high Galactic latitude of $20\deg.3$ that, for the distance quoted above, corresponds to a height of $z=426$ pc above the Galactic plane. If \psrj was born within the scale-height of massive stars in  the Galactic disk and its true age is not much longer than $\tau_c$, a projected velocity of $\sim$2800 km~s$^{-1}$ is required.
This seems unlikely, given that the average velocity of pulsars is about 500 km~s$^{-1}$ and the tail of the distribution extends no further than 2000 km~s$^{-1}$ \citep{hob05}. 
\citet{2013ApJ...778..120H} derived an upper limit of 60 mas yr$^{-1}$ for its proper motion based on \cha observations with a 10 years separation. This corresponds to a projected velocity smaller than 350 ($d/1.2$ kpc) km s$^{-1}$. 
Furthermore, diffuse X-ray emission extending  $\sim$5--7' to the South-West direction, if interpreted as a nebula trailing the pulsar, implies a motion
towards the Galactic plane \citep{2008ApJ...684..542K,2021ApJ...916..117B}. 
It is thus more likely that \psrj was born in the Galactic halo from a runaway massive star or through accretion-induced collapse of a halo population star \citep{2002ApJ...564..333M,2013ApJ...778..120H}.
Such a possibility is shared with only two other young ($\tau<1$ Myr) and distant ($d>1$ kpc) pulsars, i.e. PSR B0919$+$06 \citep{2001ApJ...550..287C} and 1RXS J141256.0$+$792204 aka Calvera \citep[and references therein]{2021ApJ...922..253M}.

\psrj was first detected in X-rays with the \cha satellite in 2001 and later observed with two \xmm pointings yielding a 67 ks exposure time in total \citep{2008ApJ...684..542K}.
Using these data, \citet{kar12c} discovered an absorption line in the phase-resolved X-ray spectrum. 
The line properties appeared to vary as a function of the pulsar spin phase, thus ruling out an instrumental/calibration effect. Phase-dependent spectral lines had been reported earlier for sources belonging to some peculiar classes of isolated neutron stars (INSs), i.e. the X-ray dim isolated neutron stars \citep[XDINSs,][] {kap08,tur09} and the central compact objects \citep[CCOs,][]{del17}. Conversely, this was the first evidence for such a feature in an ordinary RPP; lines in other RPPs were later found in PSR B1133$+$16 \citep{2018A&A...615A..73R} and PSR B0656$+$14 \citep{aru18}.
Recently, \psrj was detected also at $\gamma$-ray energies by integrating ten years of \fermi-LAT data \citep{4FGL}, with evidence of pulsations reported in \citet{2019ApJ...871...78S}.

Here we report the results of new \xmm observations of \psrj carried out in 2017-2018. Thanks to a nearly tenfold increase in the exposure time, these data represent a significant improvement compared to the 2006 observations, allowing a better characterisation of the pulsar spectral and timing properties.

\section{Observations and data analysis} \label{sec:data}

\psrj was observed four times by \xmm in 2017-2018, using the same setting for the European Photon Imaging Cameras (EPIC) instrument in all observations.
The EPIC-MOS1/2 cameras \citep{tur01} were operated in full-window mode (time resolution 2.6 s) with medium and thin optical filter, while the EPIC-pn camera \citep{str01} was in small-window mode (time resolution 5.7 ms) with the thin filter. Only the latter could be used for the timing analysis and phase-resolved spectroscopy. A log of the \xmm pointings of \psrj is given in Table \ref{tab:log}, where, for completeness, we include also the two previous observations.

The data reduction was performed using the \textsc{epproc} and \textsc{emproc} pipelines of version 18 of the Science Analysis System (SAS)\footnote{https://www.cosmos.esa.int/web/xmm-newton/sas}. 
Time intervals of high background were removed with the \textsc{espfilt} task with standard parameters, resulting in the net exposure times given in Table~\ref{tab:log}.

For the timing analysis, we selected single- and multiple-pixel   events (\textsc{pattern}$\leq$4 for the EPIC-pn and $\leq$12 for the -MOS). We converted the time of arrivals to the Solar System barycentre using the JPL DE405 ephemeris with the tool \textsc{barycen}. 
The same event selection was used for the EPIC-MOS spectral analysis, while that of the EPIC-pn was done using only single-pixel events in order to reduce the background at the lowest energies (task \textsc{espfilt} with standard parameters).

We selected source counts in a circular region of radius 30$''$, while the background was chosen far away from the pulsar and avoiding the region of the tail emission (the latter was assessed using the EPIC-MOS images, where the tail can be seen better thanks to their larger size).

The spectral analysis was performed using XSPEC (version 12.11.0). The spectra from the EPIC-pn camera were rebinned using the \textsc{grppha} tool with a minimum of 300 (phase-averaged) or 100 (phase-resolved) counts per bin, while those from the EPIC-MOS cameras with a minimum of 50 counts per bin.
The interstellar medium absorption was accounted for using the \textsc{tbabs} models with cross sections and abundances of \citet{2000ApJ...542..914W}. We give all the errors at $1\sigma$ confidence level.

\setlength{\tabcolsep}{0.6em}
\begin{table*}
\centering \caption{Journal of the \xmm observations of \psrj and PSR B1055$-$52}
\label{tab:log}

\footnotesize
\begin{tabular}{lcccccccccc}
\toprule
\midrule
Obs. ID & Start time & End time  & \multicolumn{3}{c}{Net Exposure time (ks)} & \multicolumn{3}{c}{Modes and filters} \\[3pt]
        & (UTC)        & (UTC)       & EPIC-pn & EPIC-MOS1 & EPIC-MOS2 & EPIC-pn & EPIC-MOS1 & EPIC-MOS2  \\[3pt]
\midrule
0403570101 & 2006-09-28 00:45:51 & 2006-09-28 12:14:28  & 27.60 & 38.34 & 38.67 & SW-TN & FW-TN & FW-TN\\
0403570201 & 2006-09-30 01:08:41 & 2006-09-30 08:20:33  & 17.63 & 24.30 & 24.43 & SW-TN & FW-ME & FW-ME\\
0803080201 & 2017-09-20 16:25:12 & 2017-09-22 04:50:12 & 67.73 & 88.01 & 89.06 & SW-TN & FW-TN & FW-TN\\
0803080301 & 2017-10-04 15:25:30 & 2017-10-06 04:02:10 & 76.49 & 100.70 & 101.43 & SW-TN & FW-TN & FW-TN\\
0803080401 & 2018-03-05 04:41:46 & 2018-03-06 16:46:46 & 47.69 & 64.59 & 71.43 & SW-TN & FW-TN & FW-TN\\
0803080501 & 2018-04-04 00:23:42 & 2018-04-05 15:12:03 & 76.98 & 108.39 & 118.00 & SW-TN & FW-TN & FW-TN\\
\midrule

0842820101 & 2019-06-20 18:22:24 &  2019-06-21 17:40:39 & 47.20 & 66.39  & 66.14 & SW-TN & FW-ME & FW-ME\\
0842820201 & 2019-07-09 00:17:40 &  2019-07-09 22:29:35 & 51.91 & 73.40 & 74.35 & SW-TN & FW-ME & FW-ME\\
\bottomrule\\[-5pt]
\end{tabular}

\raggedright
\end{table*}

\section{Results}\label{sec:res}

\setlength{\tabcolsep}{0.3em}
\begin{table}
\centering \caption{Ephemeris of \psrj, given in TDB units.}
\label{tab:ephem}

\begin{tabular}{lcc}
\toprule
\midrule
MJD range			& $51310-52014$			& $57838-58262$\\
Epoch zero & $51662$					& $58116$   \\
$\nu$ (Hz)	& $6.48983281353(9)$		& $6.48934759788(3)$  \\
$\nudot$ ($10^{-13}$ Hz s$^{-1}$)	& $-9.04066(9)$ & $-9.04657(9)$ \\
$\nudotdot$ ($10^{-23}$ Hz s$^{-2}$)	& $\dots$	& $-3.5(1)$  \\
Reference & \citet{2002ApJ...564..333M} & This work \\

\bottomrule\\[-5pt]
\end{tabular}
\end{table}

\subsection{Timing analysis}\label{sec:timing}

\begin{figure}
  \centering
  \includegraphics[width=1\columnwidth]{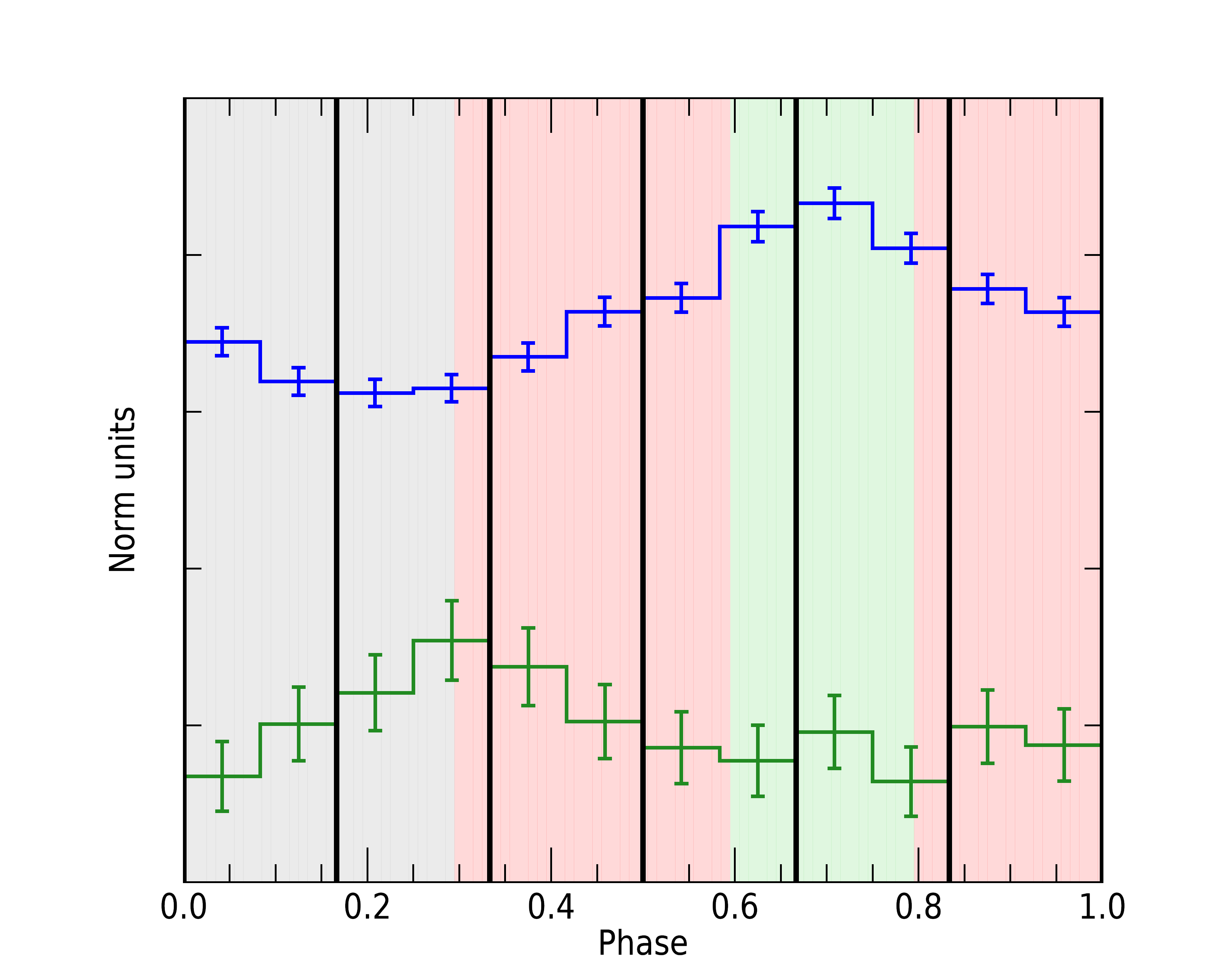}
  \caption{Pulse profiles of \psrj in the soft X-ray (0.2--1 keV, blue line) and hard X-ray (2--10 keV, green line). The black vertical lines show the six phase bins used in the phase-resolved spectroscopy (Section~\ref{sec:spectrares}). The three coloured areas correspond to the three phase bins used in Section~\ref{sec:lines}: dip (grey), rise and fall (red), and peak (green).}
  \label{fig:lc}
\end{figure}

In order to obtain the relative phase alignment of the X-ray pulse profiles with those at other wavelengths, we have phase aligned the data using a radio derived ephemeris from the Lovell Telescope at Jodrell Bank Observatory.
\psrj has been observed at a centre frequency of 1520~MHz with the Lovell Telescope since 2007, at a roughly weekly cadence, and typical rms residual of 200 $\mu$s.
We use \textsc{tempo2} to fit a timing model with spin frequency and two derivatives to data taken between 2017-03-26 and 2018-05-23, with the best-fit parameters given in Table~\ref{tab:ephem}.
For consistency between the X-ray and radio software tools, the definition of zero phase was set to MJD 58116, at infinite frequency, at the solar system barycentre, although this does not correspond to any special feature on either the radio or X-ray light curves.
For comparison, a 1520~MHz radio pulse profile from a 350 s Lovell Telescope observation on 2018-04-10 was also aligned with this ephemeris and dedispersed to infinite frequency using \textsc{psrchive}.
The pulsar exhibits significant red spin noise which prevents us from extrapolating the timing of the pulsar back to earlier X-ray observations, and so the previously published ephemeris was used for the 2006 observation, also listed in Table ~\ref{tab:ephem}. They are significantly different because of a glitch that occurred in 2012 \citep{2022MNRAS.510.4049B}.

The X-ray light curves obtained by folding all the 2017-2018 data with the radio ephemeris are shown in Figure~\ref{fig:lc}. We show the pulse profile in two energy ranges, 0.2--1 keV and 2--10 keV, and we highlight the phase bins used in the phase-resolved spectral analysis (see Sections~\ref{sec:spectrares} and~\ref{sec:lines}).

As an independent cross-check we also performed searches for pulsations in the four individual observations based  only on the X-ray data. We used the  0.3--1 keV  range, where the source is brightest and the background is negligible, and searched in the 6.4893--6.4897 Hz frequency range with the  $Z_1^2$-test \citep{buc83}.
The pulsations were significantly detected in all the observations, with best periods consistent,
within their $1\sigma$ errors, with the more precise values reported in Table~\ref{tab:ephem}. 

\begin{figure}
  \centering
  \includegraphics[width=0.49\textwidth]{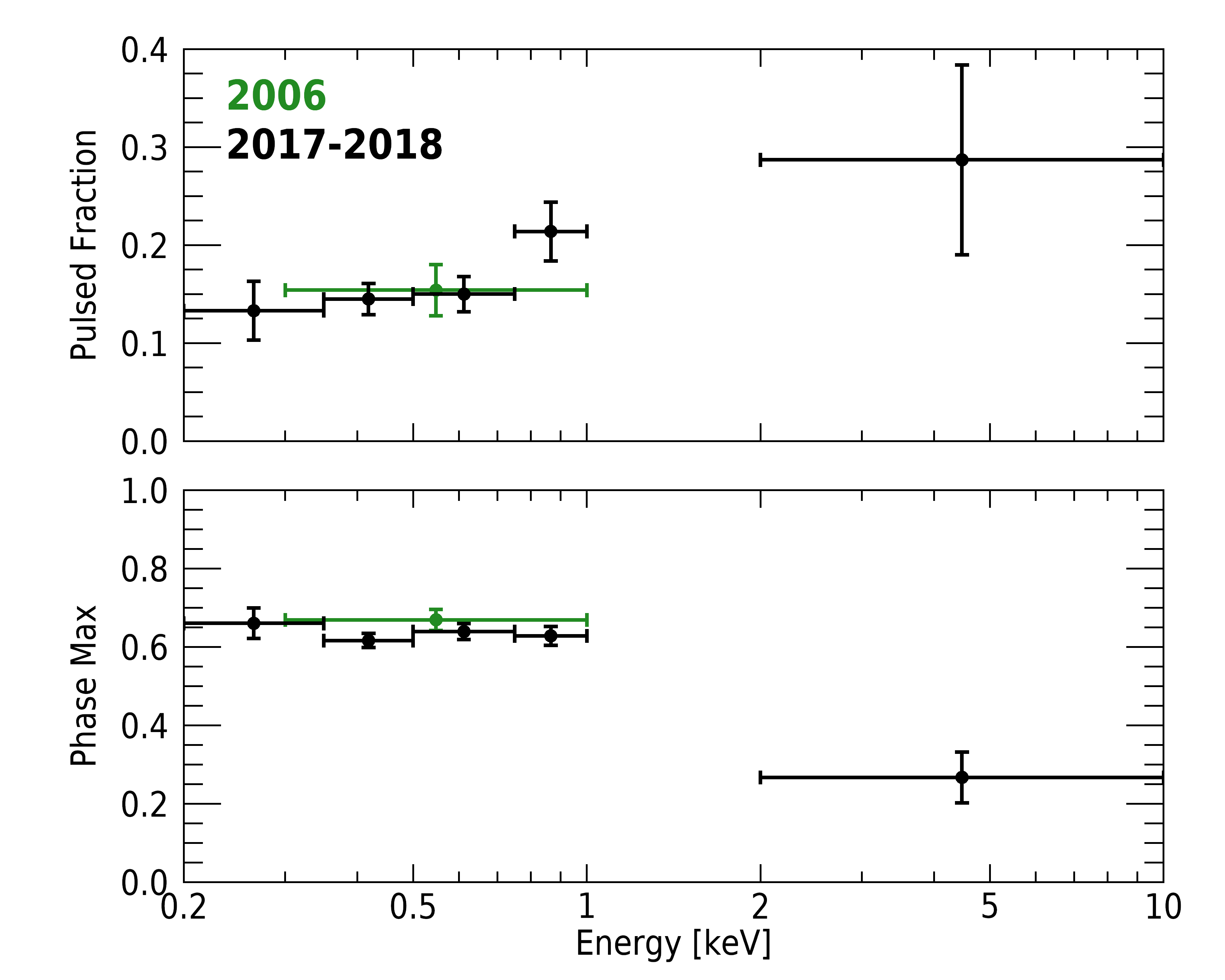}
  \caption{Pulsed fraction (upper panel) and phase of the maximum (lower panel) as a function of the energy of 2006 data (green dots) and of 2017-2018 data (black dots). Only the energy bins where the pulsation is clearly detected are plotted.}
  \label{fig:pf}
\end{figure}

To study the energy-dependence of the pulsed flux, we extracted background-subtracted pulse profiles in different energy ranges and fitted them with the function
\begin{equation}
 f(\varphi) = A + B \cos(2\pi (\varphi-\varphi_0)),
 \label{eq:cos}
\end{equation}
where $\varphi_0$ is the phase of the maximum.  We define the pulsed fraction as PF=$(A-B)/(A+B)$. The fits gave always a good description of the pulse profiles, but in the range 1--2 keV the data could be well fit also by a constant function ($\chi^2=7.60$ for 7 dof wrt $\chi^2=4.47$ for 5 dof in the case of Eq.~\ref{eq:cos}).

The derived values of the PF and $\varphi_0$, for each energy bin, are shown in Figure~\ref{fig:pf}. The PF is approximately constant below 0.75 keV ($(14.5\pm1.1)\%)$, then it increases between 0.75--1 keV ($(21.4\pm3.1)\%)$ and between 2--10 keV ($(29\pm10)\%)$. There is a difference of about $0.36 \pm 0.19$ between the phases of the maximum in the soft and hard energy ranges. The apparent lack of pulsations between 1 and 2 keV  could be due to the phase difference of the two spectral components that contribute equally in this intermediate energy range.
For comparison,  we plotted in  Figure~\ref{fig:pf} also the values that we derived from the 2006 data, where the pulsations can be seen only integrating the whole 0.3--1 keV range.

\subsection{Spectral analysis}\label{sec:spectra}

After checking that no substantial differences were present in the spectra of the four 2017-2018 observations, we combined them into one single spectrum for the EPIC-pn and one for the sum of the two EPIC-MOS. In the following all the results refer to these summed spectra.

\subsubsection{Phase-averaged spectra}\label{sec:spectraaver}

\begin{figure}
  \centering
  \includegraphics[width=0.49\textwidth]{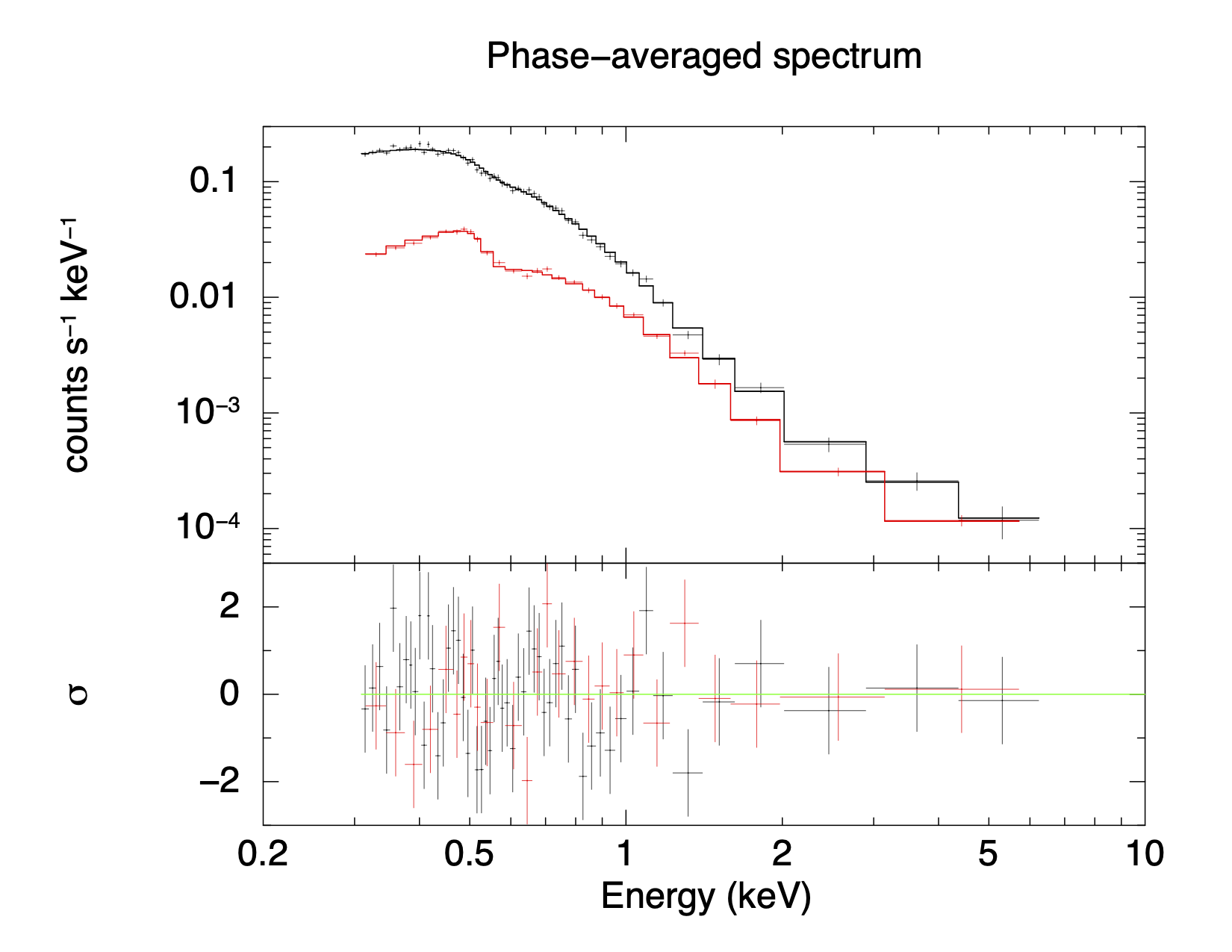}
  \caption{Phase-averaged spectrum extracted from the 2017-2018 observations with the EPIC-pn (black dots) and EPIC-MOS (red dots) cameras. The lower panels show the residuals of the best fit (2BB+PL) in units of $\sigma$.}
  \label{fig:phaaver}
\end{figure}

The EPIC-pn and -MOS spectra were fitted together in the 0.3--7 keV range with  the  model used in previous works \citep{kar12c},  which consists of three components, two blackbodies and a power law (2BB+PL), modified by interstellar absorption. We included a normalisation factor to account for possible cross-calibration uncertainties between the three cameras. 

We obtained an acceptable fit with a power-law  photon index $\Gamma = 1.80 \pm 0.17$,  blackbody temperatures $kT_1 = 70 \pm 4$ eV, $kT_2 = 137 \pm 7$ eV,  and emitting radii  $R_1 = 5.4_{-0.9}^{+1.3}$ km,  $R_2 = 0.70_{-0.13}^{+0.15}$ km (for $d=1.2$ kpc). The best-fit absorption column density is \nh $= (6.9 \pm 0.8)\times 10^{20}$ \col, and is consistent with the DM-based value (DM $=23.897\pm    0.025$, \citealt{2016A&A...591A.134B}) of $7\times 10^{20}$ \col obtained with the usual assumption of a 10\% ionization of the interstellar medium
\citep{he13}. Our measurement of \nh is also consistent with the total H\textsc{i} column density for the source position ($8\times 10^{20}$ \col) according to the sky map of \citet{2016A&A...594A.116H}.
The best-fit model is shown in Figure~\ref{fig:phaaver} and all the parameters are summarised in Table~\ref{tab:spec}.

A good fit was also found with a magnetised atmosphere model plus a power law. We used the \textsc{nsmaxg} models of XSPEC, that were computed for a partially ionized, strongly magnetised hydrogen atmosphere \citep{ho08,ho14}. We fixed $M=1.5\msun$ and $R=14$ km; the best-fit normalisation is related only to the star distance. We considered models that have a surface temperature $T\pdx{eff}$ and magnetic field $B$ distributions according to the magnetic dipole model. We explored models with $B=10^{12}$ G at the pole and angles between the line of sight and magnetic axis of 
0$\deg$ (123100) or $90\deg$ (123190). 

The first model can fit well the spectra ($\chi^2=91.21$ for 76 dof) with $T\pdx{eff}=0.35\pm0.01$ MK (that corresponds to an observed temperature $kT\approx25$ eV) and $d\approx0.25$ kpc. The second model ($\chi^2=88.29$ for 76 dof) provides a higher temperature $T\pdx{eff}\approx0.60_{-0.02}^{+0.01}$ MK ($kT\approx45$ eV) and a distance of about 1 kpc, that is more consistent with the DM-inferred value.
See Table~\ref{tab:spec} for further details.

\subsubsection{Phase-resolved spectra}\label{sec:spectrares}

\begin{figure}
  \centering
  \includegraphics[width=0.49\textwidth]{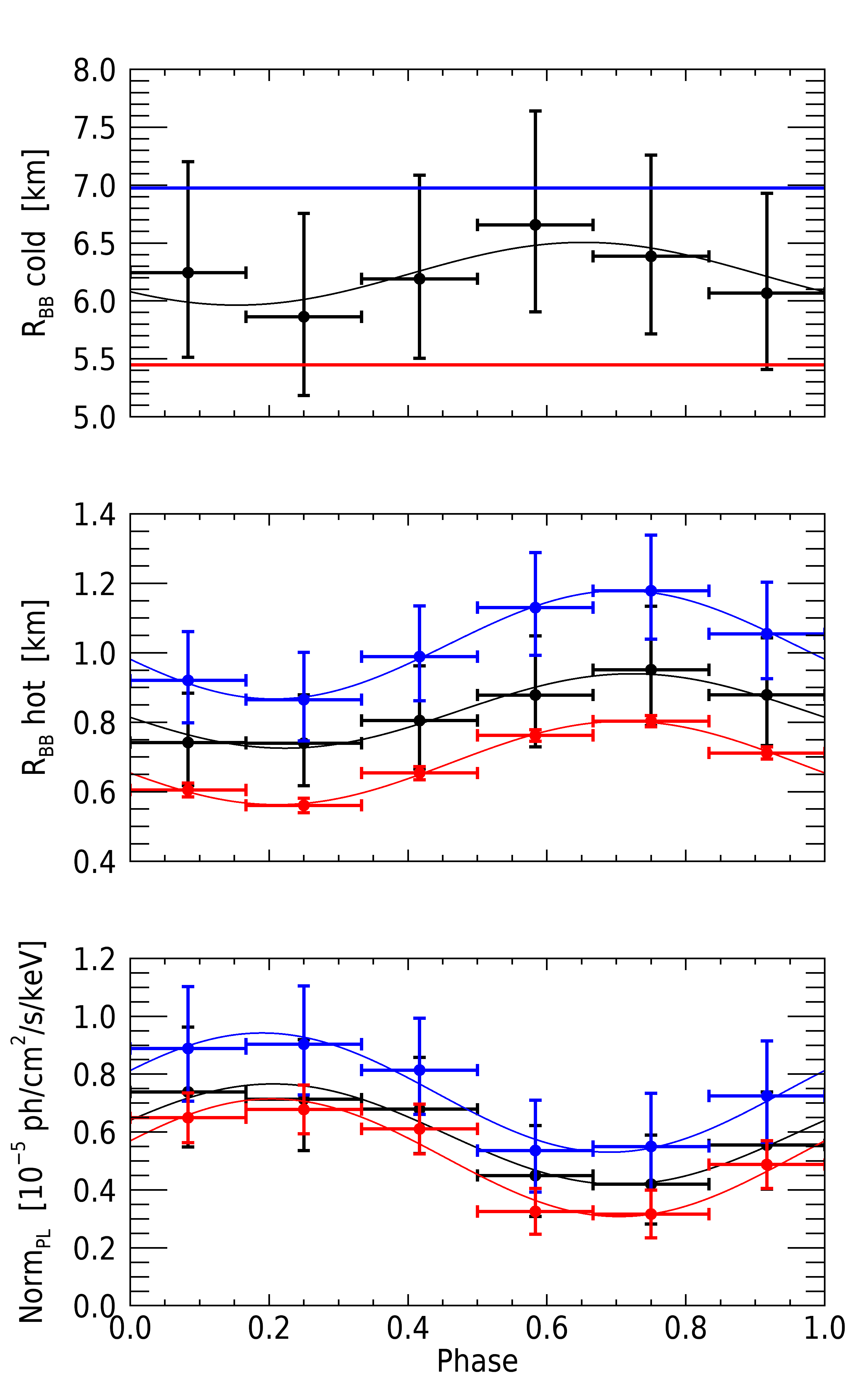}
  \caption{Best fit parameters plotted as functions of the phase: cold blackbody radius (upper panel), hot blackbody radius (middle panel) and power-law normalisation (lower panel). Black, blue and red lines correspond to the three spectral approaches listed in Table~\ref{tab:spec}, respectively. See text for further details.}
  \label{fig:phares2}
\end{figure}
 
We extracted six spectra corresponding to the six phase intervals shown by the black vertical lines in Figure~\ref{fig:lc} from the EPIC-pn data and fitted them simultaneously with the same three-components model used for the phase-averaged analysis (2BB+PL).
 
It was impossible to significantly constrain all the parameters. Therefore, we fixed the absorption to the value derived above, \nh $= 6.9 \times 10^{20}$ \col, and linked the two blackbody temperatures and the power-law index to common values for all the phases. 

In this way we obtained the best-fit parameters plotted with the black points in Figure~\ref{fig:phares2} and given in Table~\ref{tab:spec}.
A good fit was obtained, but since all the  normalisations of the colder blackbody were consistent with the same value, we imposed a common value also for this parameter. The spectral variations can be well reproduced by changing only the relative normalisations of the power law and of the hotter blackbody (blue points in Figure~\ref{fig:phares2}). 

Finally, we  tried a fit  with   the  temperatures of the two blackbodies and  the photon index fixed to the phase-averaged values. Only the normalisations of the hot blackbody and of the power law were left free to vary. Also in this case a good fit was obtained (red points in Figure~\ref{fig:phares2}).

These results indicate that the three spectral components are present at all the  phases, although with different normalisations. The phase dependence of the hotter blackbody and of the non-thermal component reflect the shape of the pulse profiles observed below 1 keV and above 2 keV, respectively. Although this spectral decomposition is clearly not unique, it involves a reasonably small number of free parameters and suggests that the colder thermal emission, with an emitting radius of $R_1=7.0_{-0.8}^{+1.1}$ km and a temperature $kT_1 = 63 \pm 3$ eV, comes from the bulk of the star's surface, while the flux from a hotter and smaller region, modulated by the pulsar rotation, accounts for the  pulsations seen at low energy. On the other hand, the pulsations at $E>2$ keV can be associated to non-thermal emission  from a different region, as indicated by phase shift between the low- and high-energy pulse profiles.

\setlength{\tabcolsep}{0.4em}
\begin{table*}
	\centering
	\footnotesize
	\caption{Spectral results}
	\begin{tabular}{l|ccccccccccc}
		\hline
		\hline
		Model & \nh & $T\pdx{CBB}$ & $R\pdx{CBB}$ & $T\pdx{HBB}$ & $R\pdx{HBB}$ & $d$ & $\Gamma$ & $N\pdx{PL}$ & $\chi^2/$ dof & dof & nhp\\
		& 10$^{20}$ cm$^{-2}$ & eV & km & eV & km & kpc &  & $\apx{a}$ & & & \\
		\hline

        Phase-averaged spectra:\\[3pt]

		2BB+PL & $6.9\pm0.8$ & $70\pm4$ & $5.4_{-0.9}^{+1.3}$ & $137\pm7$ & $0.70_{-0.13}^{+0.15}$ & $1.227~\apx{c}$ & $1.80\pm0.17$ & $5.3\pm0.9$ & 1.07 & 74 & 0.32\\[3pt]

		NSMAXG $\apx{b}$ 123100 +PL & $6.1_{-0.4}^{+0.6}$ & $25.2_{-0.8}^{+0.5}$ & $16.9~\apx{c}$ & $\dots$ & $\dots$ & $0.25_{-0.03}^{+0.02}$ & $1.90_{-0.12}^{+0.18}$ & $5.6_{-0.6}^{+1.0}$ & 1.20 & 76 & 0.11\\[3pt]

		NSMAXG $\apx{b}$ 123190 +PL & $6.4_{-0.5}^{+0.7}$ & $42.7_{-1.4}^{+0.9}$ & $16.9~\apx{c}$ & $\dots$ & $\dots$ & $0.99_{-0.12}^{+0.09}$ & $1.87_{-0.13}^{+0.17}$ & $5.4_{-0.6}^{+0.9}$ & 1.16 & 76 & 0.16\\[3pt]

		NSMAXG $\apx{b}$ 123190 +PL & $6.8_{-0.7}^{+0.4}$ & $42.2_{-1.4}^{+2.5}$ & $22.7_{-3.4}^{+2.3}$ & $\dots$ & $\dots$ & $1.227~\apx{c}$ & $1.91_{-0.16}^{+0.13}$ & $5.7_{-0.8}^{+0.7}$ & 1.07 & 76 & 0.17\\[3pt]

		\hline

		2BB+PL phase-resolved spectra: \\[3pt]
		Free norm., $kT$ and $\Gamma$ linked & $6.9~\apx{c}$ & $66\pm3$ & variable &  $130\pm7$ & variable & $1.227~\apx{c}$ & $1.9\pm0.3$ & variable & 0.92 & 173 & 0.75\\[3pt]

		$R\pdx{CBB}$ linked, $kT$ and $\Gamma$ linked & $6.9~\apx{c}$ & $63\pm3$ & $7.0_{-0.8}^{+1.1}$ &  $122\pm5$ & variable & $1.227~\apx{c}$ & $2.15\pm0.25$ & variable & 0.95 & 178 & 0.66\\[3pt]

		$R\pdx{CBB}$ linked, $kT$ and $\Gamma$ fixed & $6.9~\apx{c}$ & $70~\apx{c}$ & $5.45\pm0.05$ &  $137~\apx{c}$ & variable & $1.227~\apx{c}$ & $1.8~\apx{c}$ & variable & 0.99 & 181 & 0.53\\

  \bottomrule\\[-5pt]
\end{tabular}

\raggedright
\textbf{Notes.} Joint fits of EPIC-pn+MOS1+MOS2 phase-averaged spectra and EPIC-pn phase-resolved spectra of \psrj. Temperatures and radii are observed quantities at infinity. Errors at 1$\sigma$.
$\apx{a}$ Power-law normalization in units of $10^{-6}$ photons cm$^{-2}$ s$^{-1}$ keV$^{-1}$ at 1 keV.
$\apx{b}$ NSMAXG models \citep{ho08,ho14} with $M=1.5\msun$, $R=14$ km, a dipole distribution of the magnetic field ($B = 10^{12}$ G at the poles) and consistent temperature distribution, seen
with $\eta=0\deg$ (123100) or $\eta=90\deg$ (123190).
$\apx{c}$ Fixed value.
\label{tab:spec}
\end{table*}

\subsubsection{Absorption features}
\label{sec:lines}

No evidence for spectral lines was found in our analysis of the phase-averaged and phase-resolved spectra of \psrj when the 2BB+PL model was used.

On the other hand, \citet{kar12c} reported the presence of a phase-dependent line in the 2006 data, when fitting the continuum with a single blackbody. These authors considered three phase intervals (dip, rise and fall, peak, see Figure~\ref{fig:lc}) and found lines with  
centre at $E_0=646_{-12}^{+13}$ eV,
width $\sigma_E=137_{-13}^{+17}$ eV,
and equivalent width $EW=234\pm34$ eV
(dip),
$E_0=635_{-24}^{+11}$ eV, 
$\sigma_E=161_{-14}^{+22}$ eV,
$EW=192\pm24$ eV
(rise and fall)
and
$E_0=548\pm12$ eV,
$\sigma_E=35_{-15}^{+22}$ eV,
$EW=54\pm20$ eV
(peak).
We carried out a search for similar features in the new \xmm observation by extracting three spectra corresponding to the same phase intervals.
We fitted them in the 0.2--1.1 keV range with a single blackbody plus a Gaussian absorption line (GBB) and \nh fixed at $9.7 \times 10^{20}$ \col, as done by these authors\footnote{They limited their analysis to energies below 1.1 keV because of the poor statistics of the 2006 data, so they could neglect the contribution of the power law.}. The blackbody parameters were linked between the three spectra, while the line parameters were left free to vary.
A good fit ($\chi^2=182.92$ for 178 dof) was obtained with $kT = 93_{-5}^{+3}$ eV, $R= 7.2_{-2.1}^{+54.3}$ km and $E_0=600-650$ eV. However, the best-fit parameters for the Gaussian components are very different from those found in the 2006 data. 
We find much larger and shallower lines: $\sigma_E=343_{-16}^{+19}$ (dip), $335_{-13}^{+14}$ (rise and fall) and $337_{-20}^{+25}$ keV (peak). 
As shown by \citet{vig14}, broad absorption features with these characteristics can appear when one tries to fit with a single component  spectra well described by multiple thermal components, as is the case of \psrj. The fact that the line parameters change with the phase can be understood considering that the underlying continuum is changing. 

In order to understand if the difference between our findings and those of \citet{kar12c} are due to a long term change in the spectrum, we first repeated the same analysis on the 2006 data. With the GBB model we found  $kT = 96 \pm 2$ eV and $R=3.66_{-0.24}^{+0.28}$ km, and the spectral lines have a  $E_0=550-650$ eV and widths $\sigma_E=140_{-21}^{+28}$ (dip), $146_{-22}^{+26}$ (rise and fall) and $37_{-21}^{+23}$ eV (peak).
These results are consistent with those of \citet{kar12c}.

Then we performed simulations to estimate the probability of not revealing in the new data a feature with the properties seen in 2006 if it were still present.
To this aim we simulated $10^5$ EPIC-pn spectra for the three phase intervals with the GBB best fit parameters of the 2006 data, but with the total exposure of the  2017-2018 spectra. We fitted them with the same model, forcing the line widths to be either $<$200 eV (narrow lines) or $>$200 eV (broad lines), and compared the results computing the
maximum-likelihood ratio test \citep{1979ApJ...228..939C}.

Fitting the real data we obtained $\chi^2\pdx{narrow}=191.20$ and $\chi^2\pdx{broad}=182.94$ for 178 dof, yielding $r\pdx{obs}=0.016$.
The minimum $r$ we found after $10^5$ simulations is 36.23, more than three orders of magnitude larger than $r\pdx{obs}$.
This means that our fits of the 2017-2018 data would miss narrow absorption lines by chance is $\lesssim$10$^{-5}$. 
Extrapolation of the $r$ distribution down to the observed value shows that we would need at least $\sim$10$^7$ trials to obtain by chance $r<r\pdx{obs}$.

Finally, we wanted to address whether the \psrj spectrum changed over time, or the narrow lines detected in 2006 were statistically consistent with the 2BB+PL model seen in the new observations. We simulated $10^4$ spectra with counts distributed as the 2017-2018 2BB+PL best fit, but with the total exposure of the 2006 observations. As in the previous case, we computed $r$ by fitting the simulated spectra with GBB or 2BB models. Fitting the real 2006 data we obtained $\chi^2\pdx{GBB}=47.83$ for 60 dof, and $\chi^2\pdx{2BB}=57.63$ for 63 dof, yielding $r\pdx{obs}=134.94$. 
Based on the distribution of the $r$ values of the simulated data, this $r\pdx{obs}$ corresponds to a chance probability of $2.6\times 10^{-3}$.
This indicates that there is some evidence for a time variation of the phase-resolved spectrum.

\section{Discussion}\label{sec:disc}

\begin{figure}
  \centering
  \includegraphics[width=1\columnwidth]{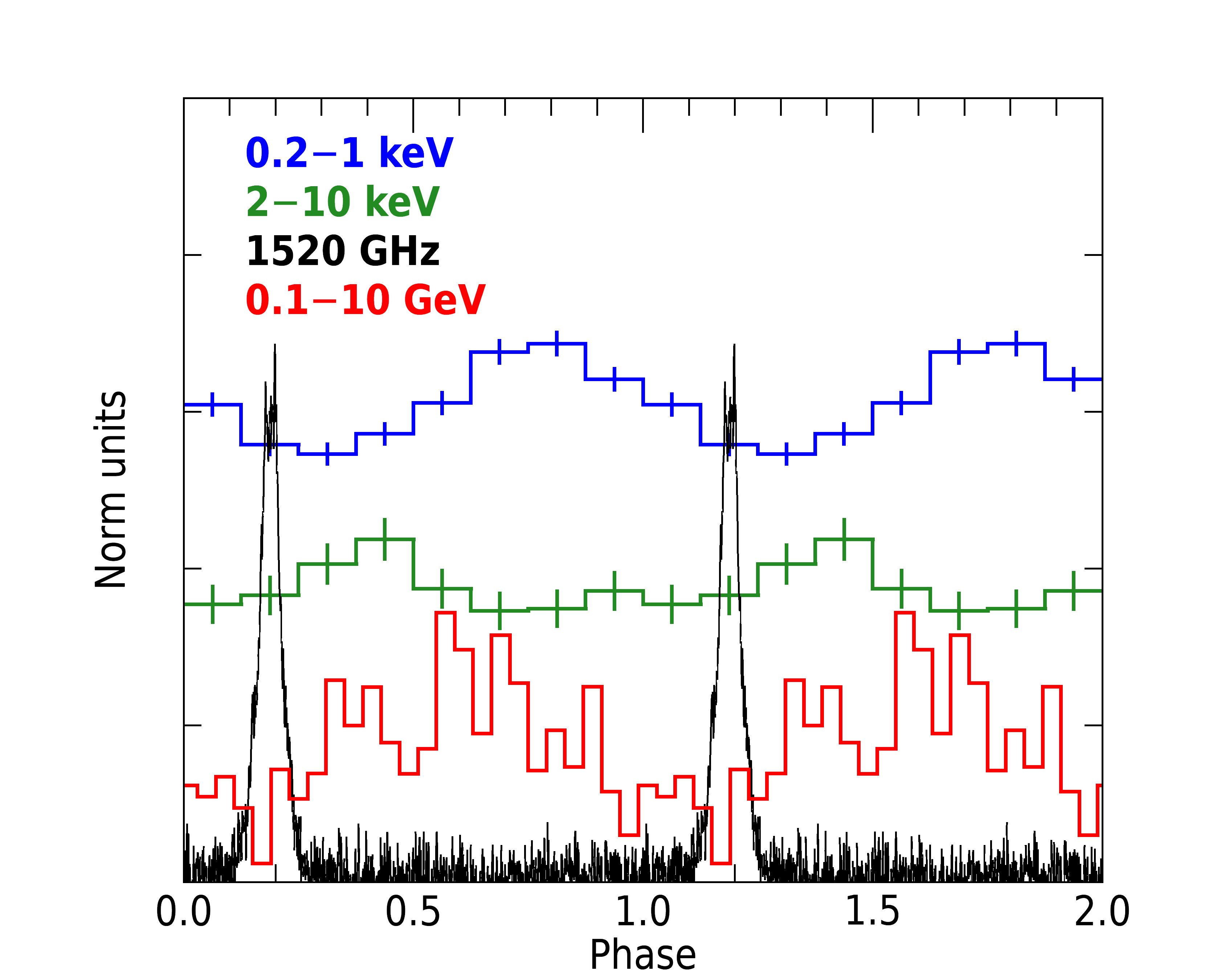}
  \caption{Multiwavelength pulsed profile of \psrj: radio (black), soft X-rays (blue), hard X-rays (green) and $\gamma$-rays (red).}
  \label{fig:multilc}
\end{figure}
The long exposure time of the new observations allowed us to better characterise the non-thermal emission of \psrj that was only poorly constrained and not seen to pulsate with the previous observations.  This component is well fit by a power law with photon index $\Gamma = 1.80 \pm 0.17$ and has a flux  $F_{1-10}=(2.44\pm0.16) \times 10^{-14}$ \flux.
The non-thermal luminosity of $(4.4\pm0.3)\times10^{30}$ \lum (for $d=1.2$ kpc), corresponds to an efficiency of about $2\times10^{-5}$, to be compared with that in the 0.1--100 GeV range of $2.8\times10^{-3}$ \citep{4FGL}.

X-ray pulsations are seen for the first time above 2 keV, with a pulsed fraction of $(29\pm10)\%$. The phase-resolved spectral analysis showed that this 
can be described with a $(40\pm10)\%$ flux modulation of a power-law component with fixed slope and  maximum flux peaking nearly in anti phase with respect to the thermal emission. 
In Figure~\ref{fig:multilc} we show also the  radio (1520 GHz) and $\gamma$-ray (0.1--10 GeV, \citealt{2019ApJ...871...78S}) pulse profiles, that have been phase-aligned using radio ephemeris we specifically derived for the epochs of the X-ray observations.
The small count statistics of the $\gamma$-ray pulse profile does not allow to draw strong conclusions on its relative alignment with that of the hard X-rays.

On the other hand, the phase difference of $0.44\pm0.01$ between the radio pulse and the maximum of the thermal emission is significant. In a simple scenario, the surface temperature is expected to peak at the magnetic poles, and the radio emission is produced within the open field lines region above the polar caps. Thus one would expect the radio and thermal emission to peak at the same phase.
The combination of field-line sweep back and aberration can introduce a phase shift between the radio emission, which comes from a significant height in the magnetosphere, and the thermal X-rays emitted at the star surface.
However, in the case of \psrj the expected  shift is only $\Delta \phi\pdx{PA}=28\deg=0.08$ in phase \citep{2011ApJ...738..114R}. This cannot account for the observed difference, thus suggesting that such a simple geometry does not apply to \psrj and/or the thermal hotter spot does not coincide with the magnetic pole and, instead, the neutron star thermal map is more complex.

\begin{figure*}
  \centering
  \includegraphics[width=1\textwidth]{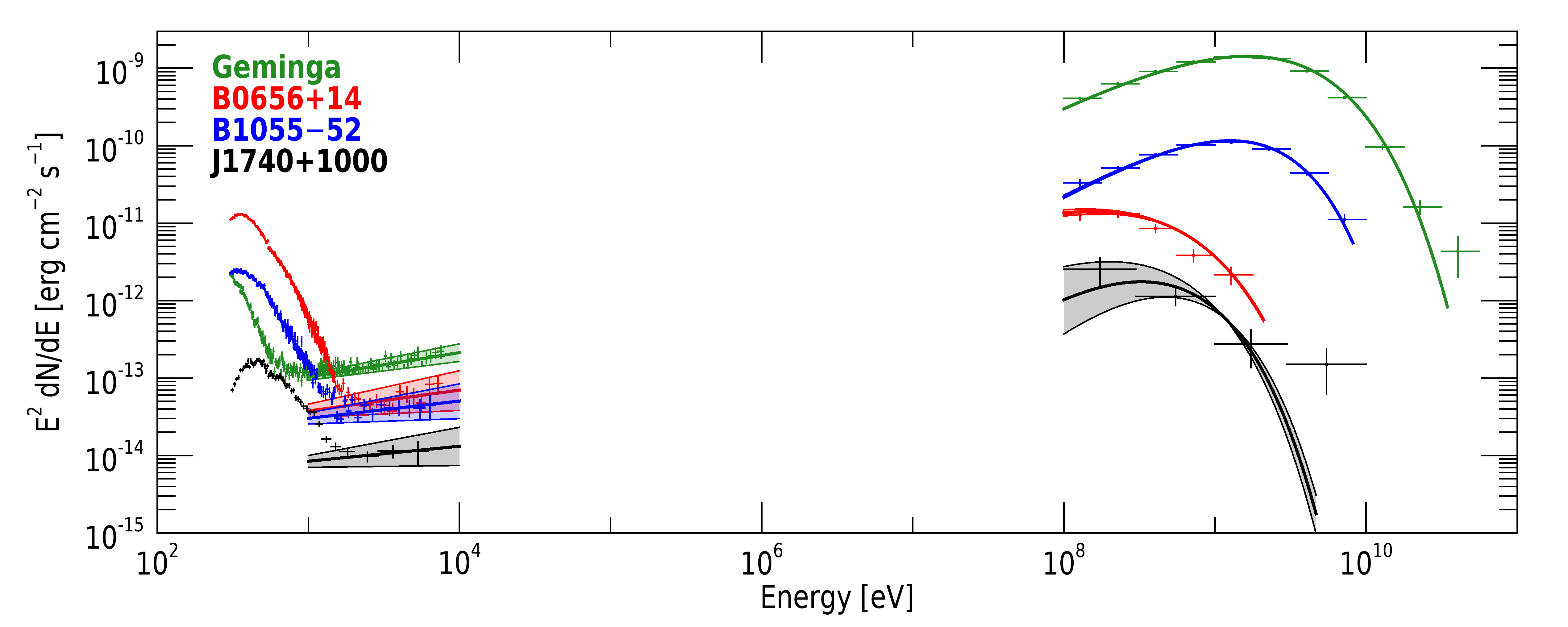}
  \caption{High-energy SED of \psrj compared to that of similar pulsars (the so called ''Three Musketeers''). The \xmm data (1--10 keV) and best fits are taken from \citet{2014ApJ...793...88M,aru18} and from  this work (see Appendix for PSR B1055$-$52). The \fermi-LAT data (0.1--100 GeV) and best fits are taken from \citet{2PC,4FGL}.}
  \label{fig:SED}
\end{figure*}

The thermal part of the X-ray spectrum of \psrj can be well fitted by the sum of two blackbodies with temperatures $kT_1 = 70 \pm 4$ eV and $kT_2 = 137 \pm 7$ eV. Phase-resolved spectroscopy revealed that  the hotter component is modulated by the pulsar rotation. The average flux corresponds to an emitting radius $R_2=683\pm8$ m. 
The cooler blackbody has an emitting radius of $R_1=5.45\pm0.05$ km, constant through the stellar rotation.

Unless the distance is much larger than that implied by the DM, the resulting emitting area of the cooler component is smaller than what is expected from the whole stellar surface. Emitting radii more consistent with the whole surface are instead obtained with atmosphere models.
In fact, the results we obtained with the \textsc{nsmaxg} model \citep{ho08,ho14} gave a radius of $\approx 14-20$ km (depending on the assumption on the star distance, see Table~\ref{tab:spec}). 

Our independent reanalysis of the 2006 observations, following the same procedure used by 
\citet{kar12c}, confirmed the presence of phase-dependent narrow lines.   
However, we could not find the same features in the 2017-2018 data that have a higher count statistics. Fitting the new spectra with the GBB model, we find that lines compatible with a reasonable fit are much larger and shallower (width of about 300 eV) than those seen in the 2006 data. As shown by \citet{vig14}, broad absorption features with these characteristics can appear when one tries to fit with a single component spectra well described by multiple thermal components, as is the case of \psrj. This led us to conclude that the broad lines observed in the 2017-2018 data are only an artifact and they just indicate a non-uniform thermal map (probably non symmetric and also responsible of the observed changes in the line parameters with phase).  

Among the $\sim$10$^5$ years old RPPs, only three objects in addition to \psrj have a 2BB+PL spectrum: these are PSR J0633+1746 (Geminga), PSR B0656+14, and PSR B1055$-$52 (the so called ``Three Musketeers",  \citealt{1997A&A...326..682B,del05}). This is due to the fact that only 10$^4$--10$^5$ years old pulsars have the cooling component visible in the X-ray band, and that long observations are needed to derive good evidence for the presence of three spectral components. 
Figure~\ref{fig:SED} shows the high-energy SEDs of the four sources. The X-ray data, obtained with \xmm, are fitted in \citet[][Geminga]{2014ApJ...793...88M}, in \citet[][PSR B0656+14]{aru18}, and in this work (see the Appendix section for PSR B1055$-$52). The \fermi-LAT data and best fits are taken from \citet{2PC,4FGL}.

The non-thermal component of the SED is fitted by a power law in the X-rays and by a cutoff power law in the $\gamma$-rays, but a self consistent unifying interpretation of broad band pulsar spectra is still lacking. \citet{2018NatAs...2..247T,2019MNRAS.489.5494T} proposed a synchro-curvature emission model that, with a small number of parameters, can describe the non-thermal spectra of thirty pulsars across seven orders of magnitude in energy. These authors did not find clear correlations between intrinsic pulsar properties and model parameters, but only hints of a relation between the magnetic field strength at the light cylinder and the parallel electric field in the magnetoshpere, that affects the peak energy in the $\gamma$-rays \citep[see also][]{2014ARA&A..52..211C}. The synchro-curvature model by \citet{2018NatAs...2..247T,2019MNRAS.489.5494T} does not allow to compute the pulse profiles and to predict the phase of the maximum. However, the different trajectories of particles that generate the SED produce a delay in the time of arrivals that can be a significant fraction of the pulsar period and can explain  the large phase differences in the X- and $\gamma$-ray pulsed profiles.

\begin{figure*}
  \centering
  \includegraphics[width=1\columnwidth]{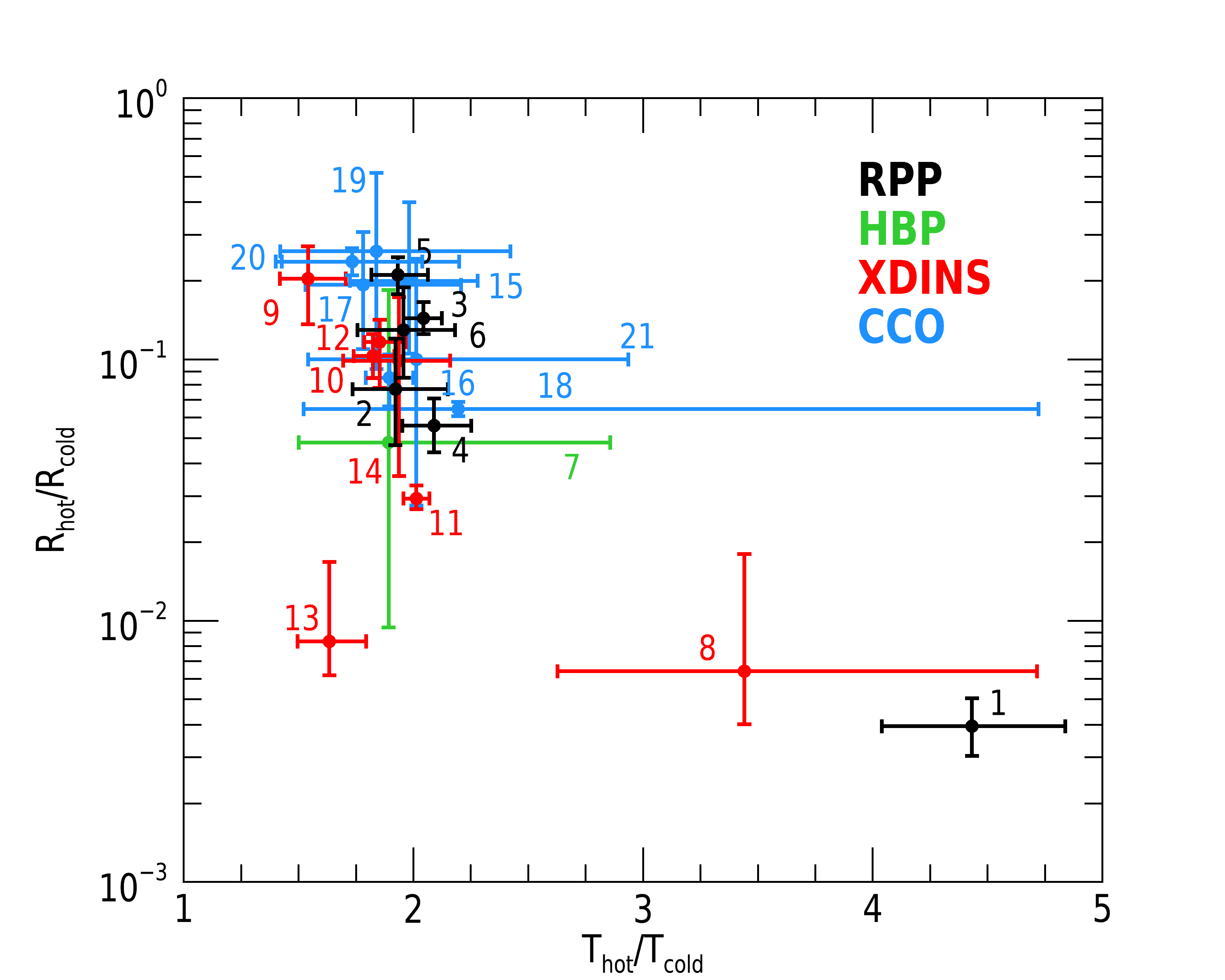}
  \includegraphics[width=1\columnwidth]{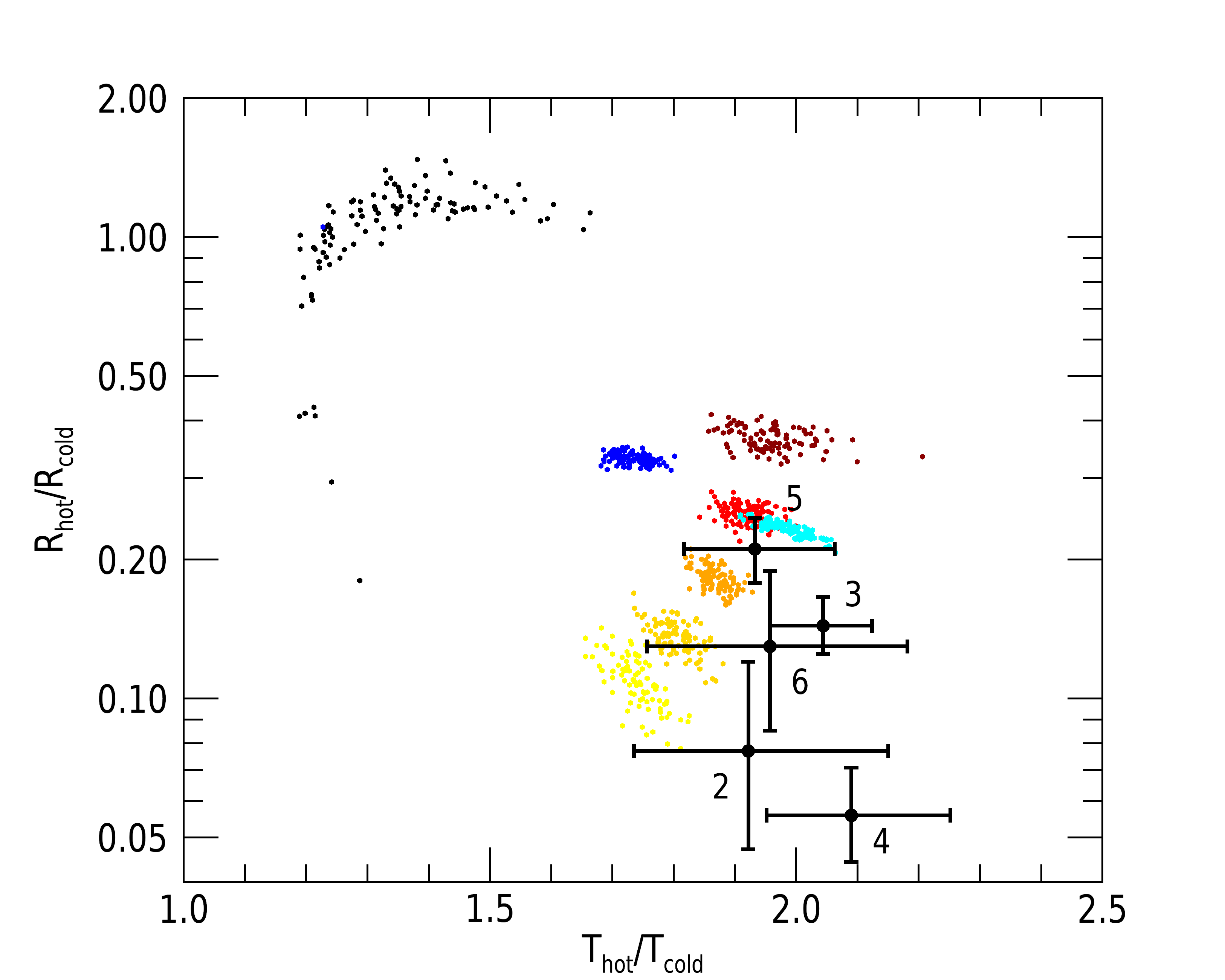}
  \caption{Ratio of the two emitting radii vs ratio of the two temperatures derived from  2BB fits of thermal spectra. Left panel: Values  observed in INSs belonging to different classes. The black points represent RPPs (\#1 Geminga, \citealt{2014ApJ...793...88M}; \#2 PSR B0656+14, \citealt{aru18}; \#3 Vela pulsar, \citealt{2007ApJ...669..570M}; \#4 PSR B1055$-$52, this work; \#5 Calvera, \citealt{2021ApJ...922..253M}; \#6 \psrj, this work), the green point represents a high-B pulsar (\#7 PSR J0726$-$2612, \citealt{2019ApJ...872...15R}), the red points represent XDINSs (\#8 RX J0420.0$-$5022, \#9 RX J0720.4$-$3125, \#10 RX J0806.4$-$4123, \#11 RX J1308.6+2127, \#12 RX J1605.3+3249, \#13 RX J1856.5$-$3754 and \#14 RX J2143.0+0654, \citealt{yon19}), the blue points represent CCOs (\#15 RX J0822.0$-$4300, \citealt{got10}; \#16 1E 1207.4$-$5209, \citealt{big03}; \#17 1WGA J1713.4$-$3949, \citealt{2004A&A...427..199C}; \#18 XMMU J172054.5$-$372652, \citealt{2011ApJ...731...70L}; \#19 XMMU J173203.3$-$344518 \citealt{2010ApJ...712..790T}; \#20 CXOU J185238.6+004020, \citealt{hal10}; \#21 CXOU J232327.9+584842, \citealt{pav09}). Right panel: Comparison between the observed values (we plot only the RPPs for clarity) and the values obtained fitting with 2BB the spectra simulated with different emission models and surface temperature distributions. The black dots represent the blackbody dipolar distribution of \citet{gre83}; the blue palette represents blackbody emission with the temperature distribution of Eq.~\ref{eq:perez} with $\alpha=64$ (blue dots) and $\alpha=4000$ (cyan dots);  the red palette represents a magnetized atmospheric dipolar distribution \citep{ho08,ho14} with $T\pdx{eff}=10^{5.5-6.3}$ K (from yellow to dark red dots). See text for details on the various models.}
  \label{fig:rapp}
\end{figure*}

In all these sources, the thermal component of the SED is well fitted by two blackbody components. In the left panel of Figure~\ref{fig:rapp} we plot in black the ratios of emitting radii and temperatures of \psrj and of the Three Musketeers,  the younger Vela pulsar \citep{2007ApJ...669..570M} and 1RXS J141256.0+792204 (``Calvera''), for which the lack of non-thermal emission is probably due to unfavourable orientation and/or large distance \citep{2021ApJ...922..253M}.
It is very striking to observe that, for all these pulsars except Geminga, the ratio $R\pdx{hot}/R\pdx{cold}$ is in the range $0.05-0.2$. Thus it is difficult to interpret the hotter and colder components in terms of emission from a  polar cap corresponding to the dipole open lines ($R\sim200-300$ m) and from  the whole surface, respectively.  Note that this conclusion is  independent of the  distance, and indicates that  the 2BB modelling is simply a first order approximation of a smooth temperature gradient across the star surface.
We also plot for comparison other classes of INSs: a high-B pulsar (green), the XDINSs (red) and the CCOs (blue). These neutron stars have very different characteristic ages (from 10$^3$ yr of the CCOs to several 10$^5$ yr of the XDINSs) and magnetic fields (from 10$^{10}$ G to 10$^{13}$ G), but with the exception of RX J0420.0$-$5022 and RX J1856.5$-$3754, they are placed in the same area of this space parameter. This means that, despite the differences, they have a similar thermal map. This is indeed quite surprising since the surface temperature is expected to evolve in time and to be sensitive to the initial magnetic field configuration in the star crust (see \citealt{2019LRCA....5....3P} for a review). Recent simulations of INS magneto-thermal evolution in 3D \citep{2020ApJ...903...40D,2021ApJ...914..118D} have shown that if the initial field is axially symmetric, although different from a simple dipole, the ensuing thermal map is itself axisymmetric. Assuming local blackbody emission, the spectrum is well fitted by a 2BB model at all times. At the inferred source age, when the star is in the so-called Hall attractor stage, the  temperatures and radii ratios are in agreement with the observed ones.  On the other hand, more complicated initial magnetic topologies result in highly asymmetric thermal maps, characterised by a higher pulse fraction, which survive over a timescale $\lesssim$100 kyr.  The behaviour shown in Figure~\ref{fig:rapp} is very striking and, in principle, may be explained by the existence of particular sets of configurations, to which the neutron star may evolve independently on its initial state, and that remain stable enough for a relatively long time,  that in turns translate into thermal maps well fitted with two thermal components in a very narrow temperature ratio (between $\sim$0.5 and $\sim$2). A systematic investigation of this scenario, carried out using systematically numerical magneto-thermal simulations, is beyond the scope of this paper, and is matter of future work. 

Instead, in this paper we tried to make a more quantitative exploration, by fitting with the 2BB model
simulated spectra previously obtained starting from different surface thermal maps and emission models.
We found that the simplest case of a blackbody emission with a temperature distribution resulting from a dipolar magnetic field  \citep{gre83} gives values of the $R\pdx{hot}/R\pdx{cold}$ and $T\pdx{hot}/T\pdx{cold}$ ratios clearly inconsistent with the observed ones (black dots in Figure~\ref{fig:rapp}, right panel). This was also pointed out by \citet{2021MNRAS.506.4593Y}. 
\citet{per06,per06b} derived an analytical approximation of the surface temperature distribution for the crust corresponding to an axially-symmetric crustal magnetic field with a strong toroidal component:
\begin{equation}
    T^4(\theta) = T^4\pdx{p} \frac{\cos^2\theta}{\cos^2\theta + \alpha \sin^2\theta} + T^4\pdx{min}
\label{eq:perez}
\end{equation}
where $\theta$ is the magnetic colatitude, $T\pdx{p}$ is the temperature at the pole, $T\pdx{min}$ is the minimum temperature reached on the surface of the star (assumed to be 0.1$T\pdx{p}$). The parameter $\alpha$ depends on the relative strength of the poloidal and toroidal components of the magnetic field and regulates the strength of the temperature gradient between the pole and the equator ($\alpha = 0.25$ corresponds to the \citealt{gre83} distribution). \citet{per06} found that the temperature distribution for a force-free magnetic field with a toroidal component present in the outer layers resembles Eq.~\ref{eq:perez} with $\alpha \approx64$.
We simulated spectra using this temperature distribution and local blackbody emissivity. When fitted with the sum of two blackbodies, in the case $\alpha=64$,  we obtain the blue dots in Figure~\ref{fig:rapp}.  Increasing $\alpha$ results in smaller values of $R\pdx{hot}/R\pdx{cold}$ and larger values of $T\pdx{hot}/T\pdx{cold}$.
To match the observed data, values of $\alpha$ larger than $\sim$4000 are needed (cyan dots), indicating a strongly concentrated region of high temperature at the poles.
On the other hand, it may be that the existence of the observed narrow ratios in the $T\pdx{hot}/T\pdx{cold}$ values is influenced by the assumption on the emission models: it is in fact well known that while using blackbody components to fit neutron stars spectral data one failed to properly account for a hardening caused by atmospheric emission. To test this effect, we used the 123190 \textsc{nsmaxg} model (see Section~\ref{sec:spectraaver}) to simulate the spectra and indeed we found a better agreement with the observations. This is shown in Figure~\ref{fig:rapp}, right panel, where we plot the radii and temperatures ratios obtained by using this atmospheric model and varying $T\pdx{eff}$ from 0.3 MK (yellow dots) to 2 MK (dark red dots). It is therefore very likely that a combination of the two effects, a relatively steep thermal map and atmospheric hardening, is responsible for the observed clustering.  

\section{Conclusions}
Our analysis of new \xmm observations of \psrj reveals that its X-ray spectrum can be well fitted by the sum of two blackbodies ($kT_1 = 70 \pm 4$ eV and $kT_2 = 137 \pm 7$ eV) and a power law of photon index $\Gamma=1.80 \pm 0.17$. Both the thermal and the non-thermal components are pulsed (PF of $(15\pm1)\%$ and $(29\pm10)\%$, respectively), and can be described by two sinusoidal functions pulsing in anti-phase. This characteristic, plus the size of the inferred emitting radii, indicate that it is difficult to interpret the hotter and colder components in terms of emission from a polar cap corresponding to the dipole open lines and from  the whole surface, respectively. It is more likely that the 2BB modelling is simply an approximation of a temperature gradient across the star surface, steeper that the one induced by a dipolar magnetic field.
This holds for both local BB emission and assuming that a magnetised atmosphere covers the star. We also found that this characteristic is shared by the majority of thermally-emitting INSs with well studied spectra, with the notable exceptions of RX J0420.0$-$5022, RX J1856.5$-$3754 and Geminga (in this respect, we note that \psrj is more similar to the other two Musketeers than Geminga).

Our independent reanalysis of the 2006 observations, with the same procedure of \citet{kar12c}, confirmed the presence
of phase-dependent narrow absorption lines. However, we could not find such features with the same profiles in the higher quality 2017-2018 data. We cannot rule out the possibility that the previous detection was a statistical fluctuation (probability of $2.6 \times 10^{-3}$). However, such rather small probability favours the possibility that the stellar properties might have slightly changed.
This could be related to the glitch that occurred in 2012 \citep{2022MNRAS.510.4049B}, between the old and the new \xmm observations.
We note also that long term variations in the soft X-ray emission of isolated NS has been observed also in RX J0720.4$-$3125 \citep{2004A&A...415L..31D,2004ApJ...609L..75V}.

Finally, we derived the multiwavelength pulse profile of \psrj. While the small count statistics of the recently detected $\gamma$-ray pulse profile \citep{2019ApJ...871...78S} does not allow to draw strong conclusions on its relative alignment with the non-thermal X-ray pulse, there is a significant misalignment ($0.44\pm0.01$ in phase) between the radio  and the thermal X-ray emission. The combination of field-line sweep back and aberration cannot explain such a phase shift, again suggesting that a simple dipole geometry does not apply to \psrj.
Large phase differences between radio and thermal X-rays have been seen also in other middle-aged \citep[e.g.][]{del05,aru18},  and old pulsars \citep{aru19}.

Our findings support the mounting evidence that  ordinary RPPs may have complicated thermal maps, as already seen for XDINSs and magnetars in quiescence. The latter were successfully interpreted with  3D magneto-thermal simulations of strong and tilted toroidal magnetic fields \citep{2020ApJ...903...40D,2021ApJ...914..118D,2021NatAs...5..145I}. It would be interesting to apply similar models also to other classes of INSs and, even more, to investigate the possible cause of the observed clustering in temperatures and radii ratios, that cover sources with a large spread of ages and magnetic field strengths. This will be matter of future study. 

\section*{Acknowledgements}
The scientific results reported in this article are based on data obtained with \xmm and the Lovell Telescope at Jodrell Bank.
The data analysis has benefited from
the software provided by the NASA's High Energy Astrophysics Science Archive Research Center (HEASARC).
We acknowledge financial support from the Italian Ministry for University and Research through grant 2017LJ39LM ``UnIAM'' and the  INAF ``Main-streams'' funding grant (DP n.43/18). 


\section*{Data availability}

All the data used in this article are available in public archives.

\bibliographystyle{mnras}
\bibliography{biblio}

\section*{Appendix\\ X-ray spectral analysis of PSR B1055$-$52}
In this section we report the spectral analysis of PSR B1055$-$52, which has two recent \xmm observations (obs. IDs 0842820101 and 0842820201, see Table~\ref{tab:log} for details). We followed the same data reduction processes described in Section~\ref{sec:data}.

We fitted the EPIC-pn and -MOS phase-averaged spectra of the two epochs in the 0.3--9 keV range with the 2BB+PL model.
We obtained an acceptable fit ($\chi^2=484.05$ for 455 dof, nhp=0.17) with  \nh $= (1.4 \pm 0.2)\times 10^{20}$ \col,  blackbody temperatures $kT_1 = 70 \pm 1$ eV, $kT_2 = 151 \pm 6$ eV,  and emitting radii  $R_1 = 4.65_{-0.24}^{+0.26}$ km,  $R_2 = 0.23\pm0.03$ km (for $d=0.350$ kpc, see \citealt{2010ApJ...720.1635M}). The power-law  photon index is $\Gamma = 1.85 \pm 0.10$, and the flux between 1--10 keV is $(8.8\pm0.3)\times 10^{-14}$ \flux.

The results are in good agreement with what found by \citet{del05}, but the analysis allowed us to plot updated spectra and spectral parameters on Figure~\ref{fig:SED} and \ref{fig:rapp}.


\bsp	
\label{lastpage}
\end{document}